\documentclass[floatfix,10pt,twocolumn,showpacs,amsmath,amssymb]{revtex4}

\usepackage{epsf}
\usepackage{graphicx}  
\usepackage{dcolumn}   
\usepackage{bm}        

\newcommand{\be}{\begin{equation}}
\newcommand{\en}{\end{equation}}
 \newcommand{\bea}{\begin{eqnarray}}
 \newcommand{\ena}{\end{eqnarray}}
  \newcommand{\sch}{Schwarzschild}

\begin{document}

\title{Schwarzschild solution as a result of thermodynamics}
\author{Hongsheng Zhang$^{1,2~}$\footnote{Electronic address: hongsheng@shnu.edu.cn}, Sean A. Hayward$^1$ \footnote{Electronic address: seanahayward@yahoo.co.uk}, Xiang-Hua Zhai$^1$ \footnote{Electronic address: zhaixh@shnu.edu.cn} and Xin-Zhou Li$^1$ \footnote{Electronic address: kychz@shnu.edu.cn} }
\affiliation{ $^1$Center for
Astrophysics, Shanghai Normal University, 100 Guilin Road,
Shanghai 200234, China\\
$^2$State Key Laboratory of Theoretical Physics, Institute of Theoretical Physics, Chinese Academy of Sciences, Beijing, 100190, China
}

\date{ \today}

\begin{abstract}
 We obtain the ~\sch~ solution from thermodynamic considerations using the assumptions of a quasi local mass form (the Misner-Sharp mass) and geometric surface gravity in a spherically symmetric spacetime. The deduction is extended to other cases such as the de Sitter, anti-de Sitter, Reissner-Nordstr$\ddot{o}$m and higher dimensional spacetimes. This paper demonstrates the simple hypotheses to obtain these known spherically symmetric solutions via thermodynamics, where essentially the Misner-Sharp mass is the mass for an adiabatic system.

\end{abstract}


\preprint{arXiv: }
 \maketitle

\section{Introduction}
   The \sch~ solution is the most important one in gravity theory, and one of the most important solutions in physics. In principle, we can consider an infinite number of effects by which general relativity (GR) could be tested. In practice, however, the gravitational effects of GR are too tiny to measure in most cases. The solar neighborhood closely approximates a spherically symmetric static space-time, and hence the solar system present a suitable arena to test GR as manifested in \sch~metric. Three classical GR effects - the anomalous perihelion precession of Mercury's orbit, the deflection of light by the Sun,  and the gravitational redshift of light, only verify the \sch~ metric. On the other hand, Newton's theory confronts serious difficulties when applied to the universe as a whole. Therefore, GR has been an expectation to provide a firm foundation in the treatment of cosmology for a long time. In some sense we can say that GR predicts a dynamical universe. However, the recent detailed observations of the dynamical evolution of the universe display a gap between theory and observations. There are still some serious problems, such as dark matter and dark energy in relativistic cosmology. That is why numerous modified theories of gravity are suggested. To sum up, the experimental verifications of GR are actually limited to the \sch~metric. We can speak in a sense that these verifications just verify \sch~ solution. Thus, it is very interesting to check the simple conditions to derive the \sch~ solution rather than the full theory of GR.

   One of the most notable achievements for the theoretical physics to take place in the 1970's was the discovery of a close relationship between laws of black hole physics and the ordinary laws of thermodynamics \cite{wald}. In fact, black hole thermodynamics is space-time thermodynamics in the sense that the physical quantities in black hole thermodynamics
    usually have global nature for a manifold. Moreover, one may be curious about the relation between the Einstein equations themselves and thermodynamic laws. In a general space-time, even the physical quantities like mass, entropy, angular momentum do not make sense since there is no Killing field, hence it is very difficult to construct thermodynamics. Furthermore, the thermodynamics in a general space-time is usually non-equilibrium state one that we know little about even for ordinary matter. However, this does not prevent us from exploring the inverse problem, i.e., to derive the Einstein equations from thermodynamic considerations. Jacobson got the Einstein equations on a hypersurface tilting to a null surface by using the local first law of equilibrium thermodynamics, under the assumptions that the entropy is proportional to the area of the local Rindler horizon of an infinitely accelerated observer and the temperature is the Hawking-Unruh temperature sensed by this observer \cite{jaco}. In fact, the concept of null observer is suggested in this research, which is a bold generalization of the concept of observer.

     The Einstein equations are 10 second-order partial differential equations which correspond to the 10 components of symmetric metric $g_{\mu \nu}(x)$. Any metric whatever is a ``solution" if no restriction is imposed on the energy-momentum tensor $T_{\mu \nu}(x)$, since the Einstein equations become just a definition of $T_{\mu \nu}$. A general metric could be of course far from equilibrium. The technical details of this problem are studied in \cite{ted2}, which show that the order of the local Killing vector is problematic. When this problem is overcome, a stationary spacetime is obtained. This confirms the previous arguments about the relation between equilibrium thermodynamics and stationary spacetime. The non-equilibrium problem is solved in part in Verlinde's entropic force  approach for the space-time thermodynamics \cite{ver}. He works in a stationary space-time, i.e., a time-like Killing vector exists. The space-time slices  are identical to each other when we shift along the integral curves of the Killing vector. It is reasonable to assume that this space-time is in equilibrium, and thus legal to apply thermodynamics for equilibrium states on such a manifold. Strictly speaking, the derived Einstein equations in this approach are just valid in stationary space-times. Also, the Newton's 2nd law and gravity law are explained by thermodynamics. But the derived Newtonian laws must be different from the classical ones, since Unruh temperature is used in the derivation, which concentrates on the 4-acceleration rather than the 3-acceleration. The resulted Newtonian laws respect Lorentz invariance, which can be treated as the Newtonian laws reformed by Sommerfeld \cite{sommer}. The entropic force approach is followed by several works \cite{several} and some problems are pointed out, for example, the position of the holographic screen may lead to negative temperature \cite{limiao1}, and the experiments by using ground based ultra-cold neutrons seem to contradict the concept of  entropic force \cite{av}. Therefore, we restrain our ambitions to a smaller project, that is, to derive the Schwarzschild solution by some ``simple" assumptions, coming or not coming from thermodynamics. In the previous derivations of GR from thermodynamics, one has to introduce a concept essentially from quantum theory, i.e., the Unruh temperature for an accelerated observer. Our derivation of \sch~in this work is strictly restricted in classical theory without any concepts from quantum theory.

    This paper is organized as follows. In section 2 we present our two conditions that can yield the \sch~ solution. In section 3 we discuss some related topics, including asymptotic de Sitter (dS), Anti-de Sitter (AdS), Reissner-Nordstr$\ddot{o}$m, and higher dimensional~\sch~ cases. The conclusion is given in section 4.

    \section{derivation of the \sch solution}

   We take a static spherical metric ansatz as follows,
   \be
   ds^2=-f(r)dt^2+h(r)dr^2+r^2d\Omega_2^2,
   \label{2dim}
   \en
   where $f(r),~h(r)$ are any functions of $r$, $\Omega_2$ denotes a unit 2-sphere, and $\frac{\partial}{\partial t}$ is the time-like Killing vector (may be not a global one). Now we consider the thermodynamics of this spacetime. Firstly, we consider the gravitational energy (mass) in a finite volume. Opposite to intuition, the gravitational mass is a very intricate problem. It is natural that gravity is associate with energy, otherwise gravity waves become senseless. But rigorous local form of gravitational energy-momentum is forbidden by equivalence principle. Therefore, its quasi-local form becomes interesting. Numerous forms for quasi-local gravitational mass have been suggested \cite{qloc}.  However, a universally accepted form is still in absence. In \cite{diff}, Bergqvist showed that various quasi-local masses are different from each other for the Kerr metric including ones given by Komar, Hawking, Penrose, Ludvigsen-Vickers, Bergqvist-Ludvigsen, and Kulkarni-Chellathurai-Dadhich. Although there are still several possibilities, the situation is comparatively simple in a spherically symmetric space. A famous form is the Misner-Sharp energy, which has been explored for various cases in numerous works\cite{ms}. In a spherically symmetric space-time described by the metric (\ref{2dim}),
   the Misner-Sharp mass inside the sphere with radius $r$ reads,
   \be
   M_{ms}=\frac{r}{2}\left(1-h^{-1}\right).
   \label{ms}
   \en
   The spherically symmetric space-time described by the metric (\ref{2dim}) with Misner-Sharp mass is called a Misner-Sharp system in this paper.
  Considering an adiabatic Misner-Sharp system, we write the first law as follows,
   \be
   \delta M_{ms}=0.
   \label{ad}
   \en
   which implies,
   \be
   h^2-h+rh'=0,
   \en
    where a prime denotes the derivative with respect to $r$.
    Solving the above equation we immediately obtain,
    \be
    h=\left(1-\frac{C}{r}\right)^{-1},
    \en
    where $C$ is an integration constant. Back substituting $h$ into the form of Misner-Sharp (\ref{ms}), we get
    \be
    C=2M_{ms}.
    \en

    Thus we get a ``half" \sch~ metric. Note that the above demonstration shows that the Misner-Sharp mass is essentially in an adiabatic system, otherwise we would obtain a wrong form had we introduced the term $TdS$ at RHS of (\ref{ad}).   We then compare the surface gravity with the generic geometric surface gravity in a spherically symmetric space-time. The geometric surface gravity given in Ref. \cite{hayward1} in  a spherically space-time is required by the unified first law. The unified first law is a significant approach in gravi-thermodynamics. The traditional black hole thermodynamics relies on the global properties of the space-time, which depend on the asymptotic behavior of a manifold. We can use the black hole thermodynamics if we know everything of a manifold. By contrast, the unified first law only cares about a patch of a manifold, which can be applied without the knowledge of the whole manifold. The quantities involved are quasi-local ones rather than global ones, and hence they can be easily applied to dynamical space-times \cite{cai}. The energy adapted to unified first law is just the Misner-Sharp mass. This is a key point to make our reasoning be a self-consistent one.

      The surface gravity is calculated as the product of the magnitude of the 4-acceleration for a particle resting at the static coordinates and $(-g_{00})^{1/2}$. In the spherically symmetric space-time, the position of a rest particle reads,
    \be
    X^{\mu}=(t,r_0,\theta_0,\phi_0),
    \en
     where $(\theta, \phi)$ are the inner coordinates of the unit 2-sphere. By definition, the 4-velocity of this particle $U_{\mu}$ reads,
     \be
     U^{\mu}=\frac{dX^{\mu}}{d\tau},
     \en
     where $\tau$ represents the proper time of the particle.
     The 4-acceleration for this rest particle can be written as,
     \be
     A^{\mu}=U^{\nu}\nabla_{\nu} U^{\mu},
     \en
     where the derivative operator $\nabla$ is compatible with the metric. We calculate the magnitude by
     \be
     a=\sqrt{|{g_{\mu\nu}A^{\mu}A^{\nu}}|},
     \en
     and hence the surface gravity $\kappa$
     \be
     \kappa=a\sqrt{-g_{00}}=a\sqrt{f}=\frac{1}{2}(fh)^{-1/2}f'.
     \label{kappa}
     \en

     The geometric surface gravity adapted to the first law reads,
     \be
     \kappa=\frac{M_{ms}}{r^2}-4\pi rw,
     \label{kapp}
     \en
     where $r$ is just the coordinate $r$ in (\ref{2dim}). $w$ is the work term,
     \be
     w=-\frac{1}{2}I^{ab}T_{ab},
     \en
      where $h$ is the two dimensional induced metric,
      \be
      I=-f(r)dt^2+h(r)dr^2,
      \en
      and $T_{ab}$ is the energy-momentum. A vacuum implies $T_{ab}=0$. We note that this definition of surface gravity is more generic and more
      reasonable than the ordinary Killing surface gravity. A nice example is given in \cite{hayward2}.

      For the vacuum, we have
       \be
       \frac{1}{2}(fh)^{-1/2}f'=\frac{M_{ms}}{r^2}.
       \label{sureq}
       \en
       Substituting equations (5) and (6) into equation (15), we obtain
       \be
       f=\left[\left(1-\frac{2M_{ms}}{r}\right)^{1/2}+D\right]^2,
       \en
       where $D$ is an integration constant. The metric should reduce to the Minkowskian one when $M_{ms}=0$, and thus we obtain
       $D=0$ since the coordinates are written in standard form and even have no free constant to adjust. We call this Minkowskian condition. The second road to get $D$ is to consider the Newtonian approximation.
        Any physical metric has to satisfy the Newtonian approximation. The Newtonian form for a spherically metric is written as usual,
      \be
      ds^2=-(1+2\phi)dt^2+(1-2\phi)dr^2+r^2d\Omega_2^2,
      \en
      where $\phi=-M/r$ is the Newtonian potential and $M$ labels the central mass. The Newtonian mass $M$ is just the Misner-Sharp mass $M_{ms}$ in a spherically symmetric case \cite{hayward0}. Expanding $f$ for large $r$, we obtain
      \be
      f=1-\frac{2M_{ms}(1+D)}{r}+2D+D^2.
      \en
       Therefore we obtain immediately,
       \be
       \phi=-\frac{2M_{ms}(1+D)}{r}+2D+D^2.
       \en
       Also we have to set $D=0$ to match the Newtonian law.

       Thus we complete the derivation
      of the ~\sch~ solution via thermodynamics considerations. The key points are an adiabatic Misner-Sharp system and the geometric surface gravity in the unified first law.

      \section{related topics}
      Slight improvements of the deduction in the above section can help us to obtain some other spherical solutions. We firstly deal with asymptotic dS/AdS space-times. The metric assumption (\ref{2dim}) and the definition of (\ref{ms}) remain the same in this case. For the first law in an adiabatic Misner-Sharp system, a pressure term should be introduced in the RHS of (3),
      \be
       \delta M_{ms}=-PdV,
       \en
      where $P$=constant (dS/AdS depends on the sign of $P$) is the pressure and $V$ is the volume in consideration. And then we obtain $h(r)$ for this asymptotic dS,
      \be
      h=\left(1-\frac{C}{r}+\frac{8\pi P}{3}r^2\right)^{-1}.
      \label{hde}
      \en
      Following the usual symbol, we define
      \be
      \Lambda=-8\pi P.
      \en
      Back instituting (\ref{hde}) into (\ref{ms}), we get the Minser-Sharp mass for asymptotic dS/AdS,
      \be
      M_{msd}=\frac{C}{2}+\frac{\Lambda}{6}r^3.
      \en
      The work term reads,
      \be
      w=\frac{1}{8\pi}\Lambda.
      \en
      In this case (\ref{kapp}) becomes,
      \be
      \kappa=\frac{C}{2r^2}-\frac{\Lambda r}{3}.
      \en
      And (\ref{kappa}) takes the same form since it does not depend on the concrete forms of $f$ and $h$.
      The equality of the surface gravity (15) becomes,
      \be
       \frac{1}{2}(fh)^{-1/2}f'=\frac{C}{2r^2}-\frac{\Lambda r}{3}.
       \label{sureq}
       \en
       Thus,
      \be
      f=\left[\left(1-\frac{C}{r}-\frac{\Lambda}{3}r^2\right)^{1/2}+D_1\right]^2,
      \en
      where $D_1$ is an integration constant. Similar to the former case, we obtain $D_1=0$ by using the Minkowskian condition under $C=0,~\Lambda=0$ or the Newtonian condition under $\Lambda=0$ and the large $r$ approximation.

      Next, we consider a solution with an electromagnetic field. We still adopt (\ref{2dim}) and (\ref{ms}) as the starting point. The first law in an adiabatic Misner-Sharp system is changed to \be
      \delta M_{ms}=\Phi dq,
      \en
      where $\Phi=q/r$ marks the electric potential and $q$ labels the charge residing at $r=0$.  Solving the above equation, we arrive at,
      \be
      h=\left(1-\frac{C}{r}+\frac{q^2}{r^2}\right)^{-1}.
      \en
      The Misner-Sharp mass $M_{msq}$ in this case is,
      \be
      M_{msq}=\frac{C}{2}-\frac{q^2}{2r},
      \en
      and the work term reads,
      \be
      w=\frac{q^2}{8\pi r^4}.
      \en
      Thus we reach
      \be
      \kappa=\frac{C}{2r^2}-\frac{q^2}{r^3}.
      \en
      Similar to the above section on the discussion of the asymptotic behavior of the metric, we obtain,
      \be
      f=1-\frac{C}{r}+\frac{q^2}{r^2}.
      \en
       Finally, we explore the higher dimensional case. For a spherically symmetric $n$-dimensional space-time,
       \be
   ds^2=-f(r)dt^2+h(r)dr^2+r^2d\Omega_{n-2}^2,
   \label{ndim}
   \en
           the Misner-Sharp mass inside the sphere with radius $r$ reads,
      \be
      M_{ms}=\frac{1}{16\pi G_n}
(n-2)\Omega_{n-2}r^{n-3}(1-h^{-1}),
      \en
      where $G_n$ denotes $n$-dimensional Newton constant, $\Omega_{n-2}$ represents the volume of a $(n-2)$-dimensional sphere. It is easy to check that the above mass degenerates to (\ref{ms}) when $n=4$. Similar to the procedure in 4-dimensional case, we reach $n$-dimensional \sch~ solution,
      \be
      ds^2=-(1-C/r^{n-3})dt^2+(1-C/r^{n-3})^{-1}dr^2+r^2d\Omega_{n-2}^2.
      \en

   \section{conclusion and discussions}
   The previous derivations of the Einstein equations from thermodynamics meet fundamental logic and practical difficulties. Generally, the symmetry of a solution of a theory is higher than that of the theory itself. We restrict our aspiration in deriving the most soundly tested solution of GR, i.e., the \sch~ solution. Under the assumption that the Misner-Sharp mass is essentially in an adiabatic system, we get a ``half" \sch~ solution. Then, by using the definition of the surface gravity in the unified first law we obtain the \sch~ solution. We also discuss the derivations about dS/AdS, Reissner-Nordstr$\ddot{o}$m, and higher dimensional cases. What is the physical significance of the fact that a mass form implies the whole information of a metric? In some sense, the quasi local mass form can be regarded as an example of holographic principle \cite{holo}, which says that the total information of $n$-dimensional space-time can be completely mapped to a $(n-1)$-dimensional hypersurface. Hence, one may not be surprised that a property of a hypersurface (the quasilocal mass) implies the property of the whole manifold (space-time metric). Furthermore, we will consider space-times with other symmetries \cite{Zhang} in further coming work.

 {\bf Acknowledgments.}
  H Zhang thanks Prof E Verlinde for discussions of the acceleration used to derive Newtonian theory in entropic force approach, and Prof S Wu and Dr X Wu for the orders of the Killing vector in Jacobson's approach. This work is supported by the Program for Professor of Special Appointment (Eastern Scholar) at Shanghai Institutions of Higher Learning, National Education Foundation of China under grant No. 200931271104, and National Natural Science Foundation of China under Grant Nos. 11075106 and 11275128.

\end{document}